\documentclass[aps,pra,twocolumn,a4paper,amsmath,amssymb,showpacs,%
superscriptaddress,floatfix,]{revtex4}

\usepackage{graphicx}
\usepackage{epstopdf}

\newcommand{\Op}[1]{\boldsymbol{\mathsf{\hat{#1}}}}

\def\openone{\leavevmode\hbox{\small1\kern-3.3pt\normalsize1}}

\begin{document}

\title{Protecting coherence in Optimal Control Theory: state dependent 
constraint approach} 
\date{\today}

\author{Jos\'{e} P. Palao}
\affiliation{Departamento de F\'{i}sica Fundamental II,
Universidad de La Laguna, La Laguna 38204, Spain}
\author{Ronnie Kosloff}
\affiliation{Institute of Chemistry and
  The Fritz Haber Research Center, 
  The Hebrew University, Jerusalem 91904, Israel}
\author{Christiane P. Koch}
\affiliation{Institut f\"ur Theoretische Physik,
  Freie Universit\"at Berlin,
  Arnimallee 14, 14195 Berlin, Germany}

\begin{abstract}
  Optimal control theory is developed for the task of obtaining a primary objective
  in a subspace
  of the Hilbert space while avoiding other subspaces of the Hilbert
  space. The primary objective can be a	 
  state-to-state transition or a unitary transformation. 
  A new optimization functional is introduced which leads to monotonic
  convergence of the  
  algorithm. This approach becomes necessary for molecular systems 
  subject to processes implying loss of coherence such as predissociation or ionization. 
  In these subspaces  controllability is hampered
  or even completely lost. Avoiding the lossy channels 
  is achieved via a functional constraint
  which depends on the state of the system at each instant in time. We outline
  the resulting new algorithm, discuss its convergence properties and
  demonstrate its functionality for the example of a state-to-state
  transition and of a unitary transformation for a model of cold Rb$_2$. 
\end{abstract}

\maketitle


\section{Introduction}
\label{sec:intro}

Coherent control utilizes the wave properties of matter to steer a
quantum dynamical process  to a desired outcome.
The source of control is interference, constructive to achieve the goal and
destructive to eliminate unwanted
consequences \cite{RZ00,SB03}. The agents  of control are external fields,
in particular electromagnetic fields. The experimental and theoretical challenge lies
in identifying these control fields.   
The present study is aimed at finding control fields which are constrained 
to limit the damage which the control field may cause to the controlled system. 

Theoretically, the control problem can be formulated as an inversion:
finding the field subject to the quantum dynamics which leads to the desired outcome.
Optimal control theory (OCT) has been developed as a tool to address this
problem \cite{rabitz1,k67}. It can be formulated starting from a
variational ansatz \cite{rabitz1} or using Krotov's method
\cite{SklarzPRA02,JosePRA03}. Recently the Krotov's method has been extended to
include a strict limitation on the spectrum of the optimized 
field  \cite{Gollub08}.

The most well studied task in OCT has been the goal of a state-to-state transition. 
Given an initial state $\psi_{ini}$ and  a closed quantum system, 
the field needs to be found which drives the system to a specific
final state $\psi_{fin}$. 
This task has been shown to be completely controllable
\cite{clark,rabitz2} if the fields are not restricted. 
Moreover the control landscape is favorable composed of flat ridges
such that the climb in the gradient direction 
will lead to one of the many possible solutions \cite{rabitz3}.

A more involved control task is to optimize the expectation value of
an operator at a final time, $\langle \Op B (t_{fin})\rangle$.
This task can be formulated in the framework of open quantum systems. The OCT approach
yields an iterative  solution to the inversion problem which is based on propagating the
system density operator $\Op \rho_S(t)$ forward in time and the target operator 
$\Op B(t)$ backward in time \cite{k131,k163,otzki}. 

The prospect of quantum computing has posed an even more complex control problem:
imposing a unitary transformation $\Op U$ on a subset of quantum states
which act as the quantum register.
The unitary transformation carries out a specific computational
task. This control task is equivalent to $N$ simultaneous
state-to-state transformations \cite{JosePRL02,JosePRA03,regina}. The
solution of the iterative set of equations  has been shown to 
become  exponentially more difficult with the size $N$ of the unitary
transformation \cite{JosePRA03}.
These findings are in accordance with a very complex control landscape \cite{rabitz4}.

A further step up in complexity is the task of imposing a unitary transformation 
under dissipative conditions. This task emerges in the quantum
governor \cite{k218}, in quantum information processing and it is a
traditional task in nuclear magnetic resonance (NMR) spectroscopy
\cite{glasser}. 

In any practical implementation the positive task of obtaining the
final goal has to be weighted by possible negative consequences. For
example control fields of high intensity can damage the system by
causing ionization or dissociation. A remedy for this problem consists
in restricting the population in certain lossy excited state
manifolds. This task has been the motivation for the development of
local control theory (LCT) \cite{k89,k94}.  LCT has  been applied
to lock unwanted electronic excitations \cite{Malinovsky97}
and recently to the problem of quantum information processing
where avoiding population loss becomes crucial \cite{Shlomo04,Sklarz06}.

However, OCT is more powerful than LCT and it is therefore desirable to
incorporate constraints describing negative consequences of the
control process into the algorithm. Such 
constraints 
depend on the state of the system at intermediate
times \cite{OhtsukiJCP04,KaiserJCP04,SerbanPRA05,KaiserCP06}. 
For example, the system can simply be restricted to remain in
an ``allowed'' or to avoid a ``forbidden'' subspace during 
its evolution. More elaborate examples include
imposing a predefined path between an initial and a final state
\cite{SerbanPRA05}, and maximizing the expectation value of a time-dependent
operator throughout the optimization time interval \cite{KaiserJCP04}.

Previous OCT studies which impose state-dependent constraints
\cite{OhtsukiJCP04,KaiserJCP04,SerbanPRA05} were performed for
state-to-state optimization and are based on 
the variational approach. 
In the present work, an optimization algorithm including
state-dependent constraints 
is obtained using the Krotov method for the state-to-state case
as well as for unitary transformations. The Krotov method offers the
advantage that the monotonic convergence of the algorithm can be
ensured by the choice of the imposed constraints
\cite{JosePRA03,ChrPRA04}. 
A brief comparison with the algorithms using the variational ansatz
\cite{OhtsukiJCP04,KaiserJCP04,SerbanPRA05}  
will be given for the state-to-state optimization.

The paper is organized as follows: 
The state-dependent constraints are formulated in
Sec.~\ref{sec:formalism}, and the resulting algorithm is
presented for optimization of state-to-state transition and of a
unitary transformation. A review of the Krotov method together with an
outline of the derivation of the equations presented in
Sec.~\ref{sec:formalism} is given in the Appendix.  
Sec.~\ref{sec:results} introduces a
model example and illustrates optimization under state-dependent
constraints for a state-to-state transition and for a unitary
transformation. Our findings are compared to related approaches in
Sec.~\ref{sec:disc}. Finally, Sec.~\ref{sec:concl} concludes.


\section{Formulation of  state-dependent constraints in the Krotov
  method} 
\label{sec:formalism}

\subsection{Optimization of a state-to-state transition}
\label{subsec:form_state2state}

The dynamics of the system is governed by the time dependent
Schr\"odinger equation
\begin{equation}\label{eq:schrodinger}
\frac{d}{dt}|\varphi(t)\rangle\,=\,-\frac{i}{\hbar}\Op{H}[\epsilon(t)]
|\varphi(t)\rangle\,,
\end{equation}
where $|\varphi(t)\rangle$ represents the state of the system at time $t$, and 
\begin{equation}\label{eq:hamiltonian}
\Op{H}[\epsilon]\,=\,\Op{H}_0\,-\,\Op{\mu}\epsilon(t)\,,
\end{equation}
is the system+control Hamiltonian. $\Op{H}_0$ denotes the field-free Hamiltonian,
$\epsilon(t)$ the semiclassical control field and $\Op{\mu}$ is a system
operator describing the coupling between system and field.

The objective of the optimization is to find a field which drives the
system from an initial state at $t=0$,
\begin{equation}
|\varphi(t=0)\rangle\,=\,|\varphi_0\rangle\,,
\end{equation}
to a target subspace at time $T$ representing the final time objective, such that
a minimum (or maximum) expectation value of the time-dependent
operator $\Op{P}(t)$ is maintained throughout the complete time interval $[0,T]$.
The target subspace at time $T$ is described by the projector
$\Op{D}$, e.g.,  $\Op{D}=|\varphi_f\rangle\langle\varphi_f|$
for a single target state.

In OCT, these requirements are formulated as a functional which depends on the
system state and the control, in such a way that an
optimal field corresponds to an extremum of the functional. 
That functional can be expressed as a sum over terms related to the 
different conditions imposed on the system evolution.

\subsubsection{The functional}

The complete functional is obtained as a sum over functionals
corresponding to the final time objective, to the state-dependent
intermediate-time objective (or constraint), and to the constraint over
the field.

The term corresponding to the objective at the final time $T$, the
actual target, can be expressed as 
\begin{equation}\label{eq:j_0}
J_0[\varphi_T,\varphi_T^*]\,=\,\lambda_0\,
\langle\varphi(T)|\Op{D}|\varphi(T)\rangle\,,
\end{equation}
where $\lambda_0$ is a real parameter, which can be negative or
positive, depending on whether
the functional is minimized or maximized during the
optimization. $[\varphi_T,\varphi_T^\dagger]$ emphasizes  
the bilinearity of the functional
with respect to the system state at time $T$. 
Other  possibilities for expressing this term exist
\cite{JosePRA03,OhtsukiJCP04}, but the resulting optimization
algorithms are very similar.

The state-dependent intermediate-time objective or constraint is 
represented by the functional,
\begin{equation}\label{eq:J_b}
J_b[\varphi,\varphi^\dagger]=\int_0^T\,g_b[\varphi,\varphi^\dagger]\,dt\,,
\end{equation}
where $g_b$ is taken to be 
\begin{equation}\label{eq:g_b}
g_b[\varphi,\varphi^\dagger]=\lambda_b\langle\varphi(t)|\Op{P}(t)|\varphi(t)\rangle\,.
\end{equation}
$\lambda_b$ is a real parameter which can be positive or
negative, as discussed later.  
More complicated dependences of  $g_b$ on the operator $\Op{P}(t)$ and
on the state $\varphi$ are conceivable. 

To obtain a closed algorithm,
the complete functional has to include a term depending on the field 
\cite{SklarzPRA02,JosePRA03}, 
\begin{equation}\label{eq:J_a}
J_a[\epsilon]=\int_0^T\,g_a[\epsilon]\,dt\,.
\end{equation}
Generally, $g_a$ can be written as 
\begin{equation}\label{eq:g_a}
g_a[\epsilon]=\lambda_a(t)\,[\epsilon(t)-\epsilon_r(t)]^2\,,
\end{equation}
where $\epsilon_r$ denotes a reference field and $\epsilon_r=0$
corresponds to the common choice of minimizing the field energy.
$J_a[\epsilon]$ represents
an intermediate-time objective, but one which does not depend on the
state of the system.

The complete functional is given by
\begin{equation}\label{eq:functional_j}
J[\varphi,\varphi^\dagger,\epsilon]\,=\,J_0[\varphi_T,\varphi_T^\dagger]\,+\,J_a[\epsilon]
\,+\,J_b[\varphi,\varphi^\dagger]\,.
\end{equation}
For simplicity, we omit the dependence of $\varphi$ and $\epsilon$ on time, except
for the final time $T$.
The optimization problem is now equivalent to the minimization or maximization of
this functional. For that purpose, the Krotov method is employed.
Since the Krotov method operates with real functions, a complete presentation of the
equations for this problem is somewhat cumbersome \cite{JosePRA03}. 
An outline of the derivation is given in the Appendix and only 
the final result is presented below.

\subsubsection{The optimization algorithm}

A guess field is denoted by $\epsilon^{(0)}(t)$ and 
the corresponding state $|\varphi^{(0)}(t)\rangle$ is
given by the evolution
Eq.~(\ref{eq:schrodinger}) with the initial 
condition $|\varphi(t=0)\rangle=|\varphi_0\rangle$.
In the Krotov method
a new field $\epsilon^{(1)}(t)$ which decreases (or increases) the functional
value is obtained by the following equations: A new ``state''
$|\chi\rangle$ is determined 
using the inhomogeneous equation
\begin{equation}\label{eq:chistate}
\frac{d}{dt}|\chi(t)\rangle\,=\,
-\frac{i}{\hbar}\,\Op{H}[\epsilon^{(0)}(t)]\,
|\chi(t)\rangle\,+\,\lambda_b\Op{P}(t)|\varphi^{(0)}(t)\rangle\,,
\end{equation}
with the ``initial'' condition
\begin{equation}\label{eq:chi_T}
|\chi(T)\rangle\,=\,-\lambda_0\,\Op{D}\,|\varphi^{(0)}(T)\rangle\,,
\end{equation}
cf. Eqs. (\ref{eq:chievolution},\ref{eq:chicondition}) of the
Appendix. It corresponds to the  common
OCT result modified by  the inhomogeneous term
$\lambda_b\Op{P}(t)|\varphi^{(0)}(t)\rangle$ 
which arises from the state-dependent constraint.
The state $|\chi\rangle$ is used to determine the new control field,
\begin{equation}\label{eq:eps1}
\epsilon^{(1)}(t)\,=\,\epsilon^{(0)}(t)\,-\,
\frac{1}{\hbar\,\lambda_a(t)}{\rm Im}
\left\{\langle\chi(t)|\Op{\mu}|\varphi^{(1)}(t)\rangle\right\}\,.
\end{equation}
cf. Eq.~(\ref{eq:newfield}) of the Appendix, where 
$\epsilon_r\equiv\epsilon^{(0)}$ was chosen. 
This is an implicit equation since
the state $|\varphi^{(1)}\rangle$ which depends on $\epsilon^{(1)}$
appears in the right-hand side of Eq.~(\ref{eq:eps1}). 
The numerical discretization of this implicit equation has
been widely discussed for the homogeneous case (see for example Ref. \cite{JosePRA03}).
The inhomogeneous term in Eq. (\ref{eq:chistate}) requires a modification of the 
time propagation method. A symmetrical propagation scheme is employed
based on the diagonalization of the Hamiltonian in the interleaved time
grid points, $t_i+\Delta t/2$.
The inhomogeneous term is evaluated as
\begin{equation}
\lambda_b\left(\frac{
\Op{P}(t_{i+1})|\varphi^{(0)}(t_{i+1})\rangle+
\Op{P}(t_{i})|\varphi^{(0)}(t_{i})\rangle
}{2}\right)\,.
\end{equation}
The iterative algorithm is constructed with $\epsilon^{(1)}$ as 
input to the next step of the iteration and the process is repeated until
the required convergence is achieved.

\subsubsection{Monotonic convergence}

The monotonic convergence of the algorithm is analyzed defining $\Delta$ as
the difference between the functional values before and after one iteration,
\begin{eqnarray}
\Delta &\equiv&J[\varphi^{(0)},\varphi^{\dagger(0)},\epsilon^{(0)}]\,-\,
J[\varphi^{(1)},\varphi^{\dagger(1)},\epsilon^{(1)}] \nonumber\\
&=&
\Delta_1+\int_0^T\,\left(\Delta_{2a}(t)+\Delta_{2b}(t)\right)\,dt\,.
\end{eqnarray}
The  terms $\Delta_j$ are derived in the Appendix 
and can be evaluated using Eqs. (\ref{eq:j_0}-\ref{eq:g_a}). This
yields 
\begin{equation}
\Delta_1\,=\,-\lambda_0\,\langle\zeta(T)|\Op{D}|\zeta(T)\rangle\,,
\end{equation}
cf. Eq. (\ref{eq:delta_1}), with the definition
\begin{equation}
|\zeta(t)\rangle=|\varphi^{(1)}(t)\rangle-|\varphi^{(0)}(t)\rangle\,.
\end{equation}
Furthermore, 
\begin{eqnarray}
\Delta_{2a}(t)&=&-g_a[\epsilon^{(1)}]+g_a[\epsilon^{(0)}]\,+\,\nonumber \\
&& \quad\left[\frac{\partial g_a}{\partial\epsilon}\right]_{(1)}\,
\left(\epsilon^{(1)}-\epsilon^{(0)}\right)\,,
\end{eqnarray}
cf. Eq. (\ref{eq:delta_2}), which yields for our choice $\epsilon_r=\epsilon^{(0)}$ 
\begin{equation}
\Delta_{2a}(t)\,=\,\lambda_a(t)\,\left(\epsilon^{(1)}(t)-\epsilon^{(0)}(t)\right)^2\,,
\end{equation}
and
\begin{equation}
\Delta_{2b}(t)\,=\,-\lambda_b\,\langle\zeta(t)|\Op{P}(t)|\zeta(t)\rangle\,,
\end{equation}
cf. Eq. (\ref{eq:delta_2}).

The algorithm  converges monotonically to a minimum (maximum) of the
functional if $\Delta\geq 0$ ($\Delta\leq 0$) in each iteration step.
A sufficient but not necessary condition consists in all $\Delta_j$
being larger (smaller) than zero. Let 
the operators $\Op{D}$ and $\Op{P}(t)$  be  positive-semidefinite.
Sufficient conditions  are then given by
\begin{equation}\label{eq:cmin}
\lambda_0\leq 0\,,\quad\lambda_b\leq 0\,,\quad\lambda_a(t)\geq 0\,.
\end{equation}
for minimization, and by
\begin{equation}\label{eq:cmax}
\lambda_0\geq 0\,,\quad\lambda_b\geq 0\,,\quad\lambda_a(t)\leq 0\,
\end{equation}
for maximization.

This result leads to some curious consequences. For example, let the
system be described by a discrete number of levels and assume  its
Hilbert space can be split into 
two subspaces, the ``allowed'' subspace, described by the projector
$\Op{P}_{allow}$, and the ``forbidden'' subspace, described by $\Op{P}_{forbid}$
($\Op{P}_{allow}+\Op{P}_{forbid}=\Op{1}$). The objective of the optimization
consists in some transition inside the allowed subspace, avoiding any
population transfer to the forbidden one. In the case of minimization
of the functional $J$, the latter
requirement can be expressed by one of the two following choices for $J_b$,
\begin{eqnarray}
(a)\;\;\Op{P}(t)&=&\Op{P}_{allow}\,,\quad\lambda_b\leq 0\,,\nonumber\\
(b)\;\;\Op{P}(t)&=&\Op{P}_{forbid}\,,\quad\lambda_b\geq 0\,.
\end{eqnarray}
In the case of maximization, the possibilities are
\begin{eqnarray}
(a)\;\;\Op{P}(t)&=&\Op{P}_{allow}\,,\hspace*{2cm}\lambda_b\geq 0\,,\nonumber\\
(b)\;\;\Op{P}(t)&=&\Op{P}_{forbid}\,,\hspace*{2cm}\lambda_b\leq 0\,.
\end{eqnarray}
In both cases,  $(a)$ and $(b)$ have the same physical meaning,
remaining in the allowed subspace, or equivalently, avoiding the
forbidden subspace. The choice $(b)$ is  more appealing in principle, since 
the inhomogeneous term of Eq. (\ref{eq:chistate}) would decrease and 
eventually become negligible when approaching an optimal
solution. However, only $(a)$ fulfills the sufficient conditions for
monotonic convergence. 

A note of caution must be made at this point. Eqs. 
(\ref{eq:cmin}) and (\ref{eq:cmax}) are sufficient but not necessary conditions.
Monotonic convergence can therefore be found for values of $\lambda$ 
not fulfilling Eqs. (\ref{eq:cmin}) and (\ref{eq:cmax}). This can happen if
the values of $\Delta_j$ compensate each other
to give a convergent total $\Delta$.
In addition, the analysis assumes an exact solution of the control equations. 
A limited accuracy of the numerical
implementation of the algorithm and a poor accuracy of the propagation
method can lead to the breakdown of the monotonic convergence \cite{SerbanPRA05}.


\subsection{Optimization of a unitary transformation}
\label{subsec:form_Utrafo}

The objective consists in
implementing a given unitary transformation $\Op{O}$,
up to a global phase,
in a given subspace $\mathcal{H}_{\Op{R}}$ of dimension $N_r$
described by the projector  
$\Op{R}$,
\begin{equation}
\Op{R}\,=\,\sum_{n=1}^{N_r}\,|n\rangle\langle n|\,.
\end{equation}
To this end, the parameter $\tau$ is defined,
\begin{equation}
\tau\,=\,{\rm Tr}\left\{\Op{O}^\dagger\,\Op{U}(T,0;\epsilon)\,\Op{R}\right\}
\,=\,\sum_{n=1}^{N_r}\,\langle\varphi_{fn}|\varphi_n(T)\rangle\,,
\end{equation}
where
\begin{eqnarray}
\Op{O}|n\rangle&=&|\varphi_{fn}\rangle\,,\nonumber\\
\Op{U}(t,0;\epsilon)|n\rangle&=&|\varphi_{n}(t)\rangle\,.
\end{eqnarray}
The modulus of $\tau$ is equal to $N_r$ when the target unitary 
transformation is implemented in the subspace $\mathcal{H}_{\Op{R}}$ by
the field $\epsilon$ \cite{JosePRA03}. 

The optimization problem is again formulated as a functional minimization
(maximization). The final time term is now defined by 
\begin{eqnarray}
J_0[\{\varphi_{Tn},\varphi_{Tn}^\dagger\}] &=&\lambda_0\,|\tau|^2 \\
&=&\lambda_0
\sum_{n=1}^{N_r}\,\langle\varphi_{fn}|\varphi_n(T)\rangle
\sum_{n'=1}^{N_r}\,\langle\varphi_{n'}(T)|\varphi_{fn'}\rangle\,,\nonumber
\end{eqnarray}
where 
$\{\varphi_{n},\varphi_{n}^\dagger\}$ denote the set of states $|\varphi_n\rangle$
($n=1,\dots,N_r$).
Other choices of $J_0$ are possible \cite{JosePRA03}.
The intermediate-time state-dependent term takes the form,
\begin{eqnarray}
g_b[\{\varphi_n,\varphi_n^\dagger\}]&=&\lambda_b\,{\rm Tr}
\left\{\Op{U}(t,0;\epsilon)^\dagger\Op{P}(t)\Op{U}(t,0;\epsilon)\Op{R}
\right\} \nonumber \\ &=& 
\lambda_b\,\sum_{n=1}^{N_r}\,\langle\varphi_{n}(t)|\Op{P}(t)|\varphi_n(t)\rangle\,.
\end{eqnarray}
The constraint over the field is taken to be the same as in the
state-to-state case, cf. Eq.~(\ref{eq:g_a}).


The equations defining the algorithm are obtained using the Krotov method
as outlined in the Appendix. They read as follows:
$N_r$ ``states'' $|\chi\rangle$ are given by the inhomogeneous evolution 
equation,  
\begin{equation}
\frac{d}{dt}|\chi_n(t)\rangle\,=\,
-\frac{i}{\hbar}\,\Op{H}[\epsilon^{(0)}(t)]\,
|\chi_n(t)\rangle\,+\,\lambda_b\Op{P}|\varphi_n^{(0)}(t)\rangle\,,
\end{equation}
with the ``initial'' condition
\begin{eqnarray}
|\chi_n(T)\rangle&=&-\lambda_0\,{\rm Tr}\left\{\Op{O}^\dagger
\Op{U}(T,0;\epsilon^{(0)})\Op{R}\right\}|\varphi_{fn}\rangle \nonumber
\\ &=&
-\lambda_0\,\left(\sum_{n'=1}^{N_r}\,\langle\varphi_{fn'}|
\varphi_{n'}^{(0)}(T)\rangle\right)
|\varphi_{fn}\rangle\,,
\end{eqnarray}
$n=1,\dots,N_r$. They are used to determine the field $\epsilon^{(1)}$
by means of 
\begin{eqnarray}
\epsilon^{(1)}(t)&=& \\ \nonumber
&&\epsilon^{(0)}(t)
\,-\,\frac{1}{\hbar\,\lambda_a(t)}{\rm Im}
\left\{\sum_{n=1}^{N_r}\langle\chi_n(t)|\Op{\mu}|\varphi_n^{(1)}(t)\rangle\right\}\,.
\end{eqnarray}

The discussion of the monotony of convergence  is equivalent to 
the state-to-state case \cite{JosePRA03}. 
That is, the sufficient conditions for minimization,
Eq. (\ref{eq:cmin}), and maximization, Eq. (\ref{eq:cmax}) are also
valid for the optimization of a unitary transformation. 

\section{Illustration of the algorithm enforcing a state-dependent constraint} 
\label{sec:results}

\begin{figure}[tb]\centering
\includegraphics[width=0.9\linewidth]{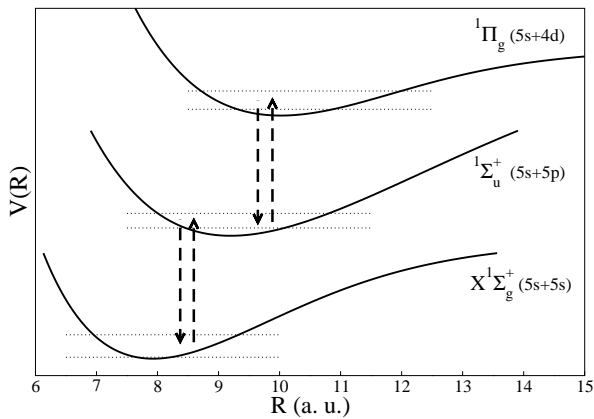}
\caption{Schematic representation of the Rb$_2$ model used in the calculations. The
dotted lines indicate the position of the vibrational manifolds
considered in each electronic state.} 
\label{fig:newpotentials}
\end{figure}

In order to illustrate the algorithm outlined in the previous section, 
a simplified model of the vibrations in a Rb$_2$ molecule is employed where
only three electronic states are considered, cf. fig. \ref{fig:newpotentials}. 
Of  each electronic state, $11$ vibrational levels were chosen, specifically
$v=0,\dots,10$ from the $X^1\Sigma_g^+$ electronic ground state,
$v'=5,\dots,15$ from the $^1\Sigma_u^+$ excited state, 
and $v''=2,\dots,12$ from the $^1\Pi_g$ excited state. 
A laser field, $\epsilon(t)$, couples the
 $X^1\Sigma_g^+$ levels to the $^1\Sigma_u^+$ 
levels and the $^1\Sigma_u^+$  to  the $^1\Pi_g$ levels.
In this model the transitions $v=0$ to $v'=10$
and $v'=10$ to $v''=6$ have
similar frequencies 
and Franck-Condon factors (FCF), 
$\omega_{1\to 10}=0.0507\,$a.u. vs. $\omega_{10\to 6}=0.0506\,$a.u., and
modulus of FCF 0.17 vs. 0.23.
The Hamiltonian describing our model is then given by
\begin{eqnarray}
  \label{eq:H}
\Op{H}&\,=\,& \sum_{i=1}^{3}\Op{H}_i\otimes |e_i\rangle\langle e_i| + \\
&& \Op{\mu}\,\epsilon(t)\,\big(|e_1\rangle\langle e_2|+|e_2\rangle\langle e_1|+
|e_2\rangle\langle e_3|+|e_3\rangle\langle
e_2|\big)\,, \nonumber
\end{eqnarray}
where a dipole moment operator $\Op{\mu}$ independent of the internuclear
distance $R$ is assumed. The vibrational Hamiltonian is denoted by
$\Op{H}_i$ and 
the electronic state associated to $X^1\Sigma_g^+$, $^1\Sigma_u^+$
and $^1\Pi_g$ by $|e_i\rangle$, ($i=1,2,3$), respectively.
The vibrational level energies  and FCF were obtained by
diagonalization of the vibrational Hamiltonians employing the
potential energy curves of Ref.~\cite{ParkJMS01}. 

For the optimization examples described below, a guess field
of the form
\begin{equation}
\epsilon_g(t)\,=\,\epsilon_0\,s(t)\,\cos[\Omega(t-T)/2]
\end{equation}
is employed. $T$ corresponds to the target time for the optimization, set
to $T=8\,$ps, and  $s(t)=\exp[-32(t/T-1/2)^2]$ is taken to be a
Gaussian shape function. The central 
frequency of the guess field is chosen to be $\Omega=\omega_{v=0\to v'=10}$ 
and $\epsilon_0=10^{-4}\,$a.u..
The optimization parameters related to the final time
objective, $J_0$ and to the field constraint, $J_a$, are set to
$\lambda_0=-1$ and $\lambda_a(t)=100/s(t)$ in all calculations.

The state-dependent constraint in the optimization forces
the population to remain in the  subspace of the two lower electronic
states. This is formulated by identifying $\Op{P}(t)\equiv\Op{P}_{allow}$ 
as the projector onto the $X^1\Sigma_g^+$ and $^1\Sigma_u^+$ levels, 
$\Op{P}_{forbid}$ corresponds thus to the projector onto the $^1\Pi_g$ levels.
This choice of the allowed subspace is motivated as follows. The
pulse duration is much shorter than the spontaneous emission
lifetimes. On the timescale of the pulse, losses in 
a molecular system are due to processes such as predissociation,
auto- or multi-photon ionization. Unlike spontaneous
emission, these processes are relevant only in certain excited electronic
states. Electronic states which are not affected by loss can therefore
be included in the allowed subspace. 


\subsection{State-to-state optimization under state-dependent constraints} 
\label{subsec:results_state2state}

The objective for state-to-state optimization  is chosen to transfer  
population initially in level $v=0$ of the electronic
ground state to  level $v=1$  at time $T$, using
Raman-like transitions via levels $v'$ in the $^1\Sigma_u^+$ excited state.
Optimizations with and without state-dependent constraint are
compared. In both cases, the optimal field achieves a 
population transfer larger than $99.9$\% at the final time  $T$.
However, without the state-dependent constraint the optimal field does
not ``know'' that it is not supposed to populate levels in the upper
electronic $^1\Pi_g$ state at intermediate times. Due to similar
transition frequencies and Franck-Condon factors, population transfer
into the forbidden subspace therefore occurs.

In order to compare the performance of optimization with and without
state-dependent constraint quantitatively, two measures are defined,
\begin{equation}
J_{norm}=\frac{J}{\lambda_0+\lambda_bT}\,,
\end{equation}
and
\begin{equation}
I_P=\frac{J_b}{\lambda_b T}=
\frac{1}{T}\int_0^T\langle\varphi(t)|\Op{P}_{allow}|\varphi(t)\rangle dt\,.
\end{equation}
Unlike the original functional Eq. (\ref{eq:functional_j}),
the normalized functional $J_{norm}$ has an optimal value equal to one 
independent of 
the choice of $\lambda_0$ and $\lambda_b$,
while $I_P$ corresponds to the average value of population in the allowed subspace. 
\begin{figure}[tb]\centering
\includegraphics[width=0.9\linewidth]{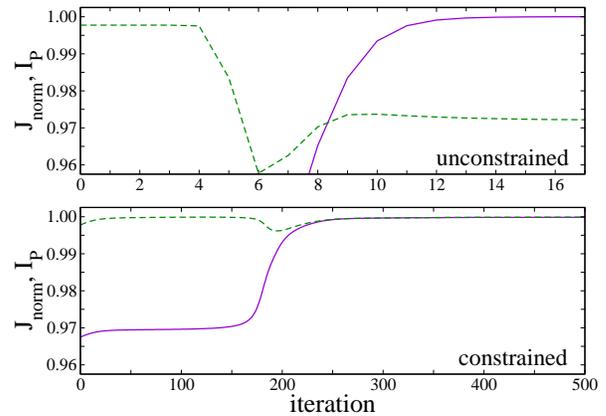}
\caption{(Color online)
  $J_{norm}$ (violet solid line) and $I_P$ (green dashed line) as a function of the 
  number of iterations for $\lambda_b=0$ corresponding to optimization
  without state-dependent constraint (upper panel) and
  $\lambda_bT=-32$ (lower panel). Note the different scale of the
  $x$-axes in the two panels.}
\label{fig:convergence}
\end{figure}
 Figure~\ref{fig:convergence} shows the values of $J_{norm}$ and $I_P$ as
the iterative optimization proceeds. 
Optimization with state-dependent constraint 
requires a larger number of iterations to reach the optimal
value of the total objective $J$. However, this price is paid off
since in this case the solution 
keeps indeed nearly all of
the population in the allowed subspace at any time.

\begin{figure}[tb]\centering
\includegraphics[width=0.9\linewidth]{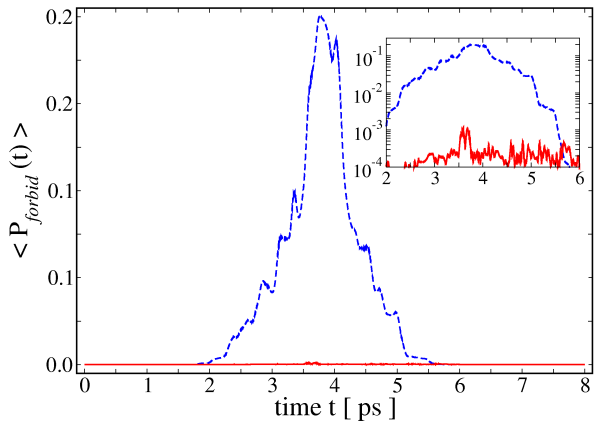}
\caption{(Color online)
  Population in the forbidden subspace as a function of 
  time for the system driven by the optimal field obtained
  with $\lambda_b=0$ after $17$ iterations (blue dashed line) and with
  $\lambda_bT=-32$ after $500$ iterations (red solid line).
  The inset shows the population in the central time interval in a
  semi-logarithmic plot.}
\label{fig:allow}
\end{figure}
\begin{figure}[tb]\centering
\includegraphics[width=0.9\linewidth]{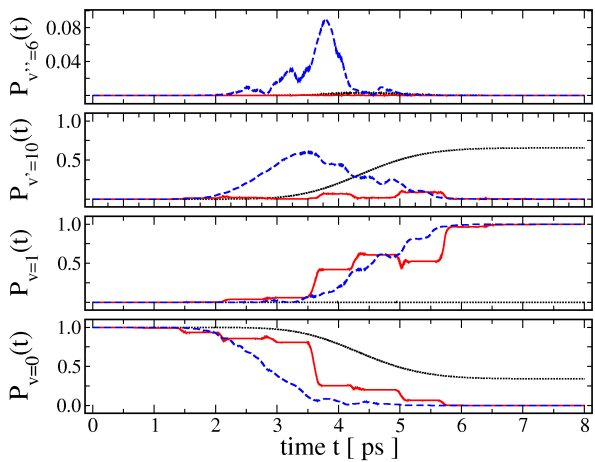}
\caption{(Color online)
  Population  in levels $v=0$, $v=1$, $v'=10$ (in the allowed
  subspace) and $v''=6$ (in the forbidden subspace) 
  as a function of time for the system driven by the optimal field obtained 
  with $\lambda_b=0$ after $17$ iterations (blue dashed line) 
  and with $\lambda_bT=-32$ after $500$ iterations (red solid line). Also
  shown is the population for 
  the evolution with the guess field (black dotted line). Note the 
  different scale of the $y$-axis for $v''=6$.}
\label{fig:population}
\end{figure}
Figures \ref{fig:allow} and \ref{fig:population} demonstrate
the behavior of the populations under the dynamics generated
by  the optimized pulses obtained with and without the state-dependent
constraint. 
The amount of population in the upper electronic
state is largely reduced for a field obtained with state-dependent
constraint as compared to that resulting from unconstrained
optimization, cf. fig. \ref{fig:allow}. 
Two different transfer mechanisms are found: The pulse obtained with
the state-dependent constraint transfers population to
$v=1$ in a ladder-like process which is driven by short
subpulses. In between the subpulses, the 
amount of population in the $^1\Sigma_u^+$ excited state (level
$v'=10$) is small, cf. fig. \ref{fig:population}. Further
excitation to the forbidden  $^1\Pi_g$ levels becomes thus unlikely. 
On the other hand, the pulse obtained without  the state-dependent constraint
transfers a large amount of population to the 
intermediate $^1\Sigma_u^+$ electronic state (mainly to level $v'=10$)
which is later to be brought back to the ground state to 
level $v=1$. The large amount of population which resides in the $^1\Sigma_u^+$
electronic state while the field is on 
allows for transient transfer  to the upper electronic state at
intermediate times.

\begin{figure}[tb]\centering
\includegraphics[width=0.9\linewidth]{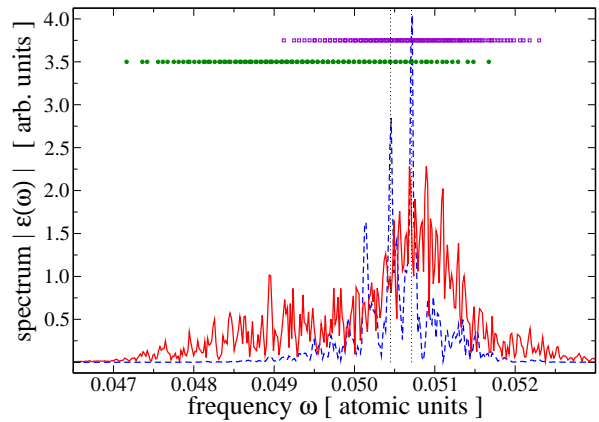}
\caption{(Color online)
  Spectral amplitude  as a function of the frequency $\omega$ for
  of the optimized field obtained  with $\lambda_b=0$ after
  $17$ iterations (blue dashed line) and with $\lambda_bT=-32$ after
  $500$ iterations    (red solid line). 
  The dashed vertical lines indicate the transition frequencies
  $\omega_{v=0\to v'=10}$ and
  $\omega_{v=1\to v'=10}$. The transition frequencies between the
  levels of the $X^1\Sigma_g^+$ electronic ground state and the
  $^1\Sigma_u^+$ intermediate state are represented by green circles,
  and the transition frequencies between the $^1\Sigma_u^+$
  intermediate state and the $^1\Pi_g$ upper state by violet
  squares.
}
\label{fig:optimal_field}
\end{figure}

Both transfer mechanisms share the common feature of a process driven
mainly by one-photon transitions.
Population is transferred to the intermediate 
electronic state by a one-photon absorption. This population is later
sent back to the electronic ground state by a one-photon emission or 
further excited to the upper electronic state by another one-photon
absorption. Large spectral amplitudes of the optimal fields 
at frequencies corresponding to the main transition frequencies of the
system reflect this finding, cf. fig.
\ref{fig:optimal_field}. Analysis of the two-photon  
spectrum, i.e. of the Fourier transform of $\epsilon(t)\epsilon(t)$, 
confirms that the amount of processes involving two or more photons is
very small.
Figure \ref{fig:optimal_field} furthermore illustrates an important
difference between the results of optimization with and without
state-dependent constraint: The additional requirement
implies a more complex optimal solution which is reflected both in a broader 
spectrum and in a more intricate dependence  of the spectrum on frequency.


\subsection{Optimization of a unitary transformation under state-dependent
  constraints}  
\label{subsec:results_Utrafo}

\begin{figure}[tb]\centering
\includegraphics[width=0.9\linewidth]{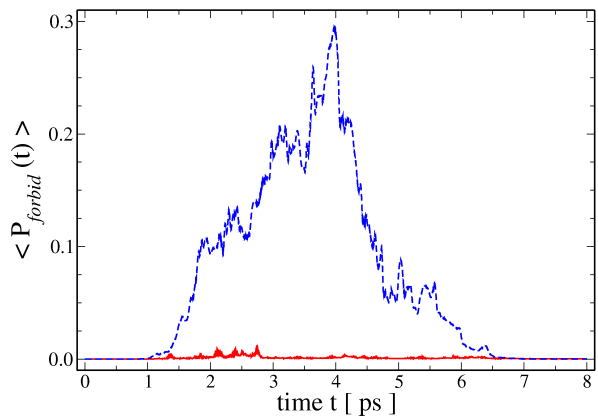}
\caption{(Color online)
  Population in the forbidden subspace as a 
  function of time for the system driven by the optimal field obtained with
  $\lambda_b=0$ after $50$ iterations (blue dashed line) and with
  $\lambda_bT=-8$ after $500$ 
  iterations (red solid line) when the system is initially in level $v=0$.}
\label{fig:ut_allow}
\end{figure}

The implementation of a Fourier transform \cite{WeinsteinPRL01}
in levels $v=0,1,2,3$ of the X$^1\Sigma_g^+$ electronic ground state
is chosen as example
objective for the optimization of a unitary transformation.
The state-dependent constraint is taken to be identical to the
optimization of state-to-state transfer, i.e. the $^1\Pi_g$ upper electronic
state represents the forbidden subspace.
Figure \ref{fig:ut_allow} shows the population in the levels of the
forbidden subspace for
the system driven by an optimized field obtained with and without 
state-dependent constraint. In both cases $|\tau|>3.999$; since the target
value is equal to $4$, the number of levels in which the unitary
transformation is implemented, this corresponds to an error of less 
than $10^{-3}$. The main results are similar to those of optimizing a
state-to-state transition: A larger
number of iterations is needed to obtain similar performance with
respect to the final
time objective $J_0$. In addition the solution becomes more complex for  
the optimization with the state-dependent constraint. 
The two tasks of a state-to-state transition and of a unitary transformation
 are run with the same number of iterations for optimization with the
 state-dependent constraint.
As expected the state-to-state transition converges faster, cf.
figs. \ref{fig:allow} and \ref{fig:ut_allow}. The effort for the
optimization of a unitary transformation is approximately equivalent to
$N$ simultaneous state-to-state transitions (here $N=4$), i.e. it
corresponds to a
more difficult optimization problem \cite{JosePRA03}.
%

\subsection{Robustness with respect to decay}

\begin{figure}[tb]\centering
\includegraphics[width=0.9\linewidth]{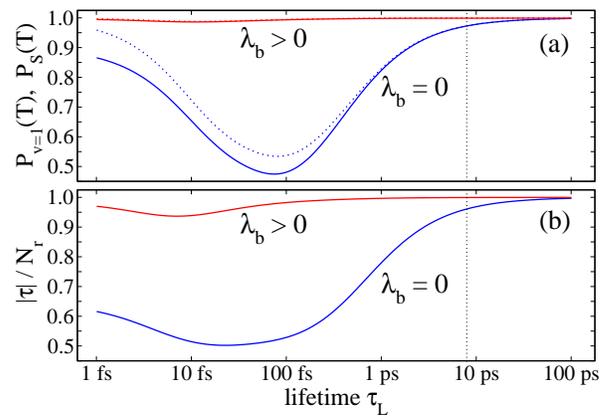}
\caption{(Color online)
  Comparison of final results at time $t=T$ as a function
  of the lifetime of the upper excited state.
  (a) State-to-state transition: Population in level $v=1$, $P_{v=1}(T)$, (solid
  lines) and total population remaining in the system, $P_s(T)$,
  (dotted lines)  for the optimized field obtained 
  with (red) and without the state-dependent constraint
  (blue). (b) Unitary transformation:
  Normalized final time objective, $|\tau|/N_r$ for a field
  obtained by optimization of the 
  unitary transformation  with (red solid line)
  and without state-dependent constraint
  (blue solid line). The vertical dotted line indicates the overall
  pulse duration.}
\label{fig:loss}
\end{figure}

A loss mechanism in the $^1\Pi_g$ electronic state is modeled by
adding an imaginary term  $-i\Gamma/2$ to the vibrational energies,
where $\Gamma=1/\tau_L$ denotes the decay rate and  
$\tau_L$ the lifetime. Physically, such a decay is caused by
processes such as predissociation or auto-ionization.
When decay is included, the system dynamics generated by the
optimized field is perturbed depending on the decay
rate and on the amount of population in
the lossy upper electronic state. 
Figure~\ref{fig:loss}a shows  
the population in the target level for the state-to-state transition with 
and without state-dependent constraint with optimization parameters
as described above (solid lines). Moreover, the total amount of
population  remaining  in the system at time $T$, 
\begin{equation}
P_s(T)=\langle\varphi(T)|(\Op{P}_ {allow}+\Op{P}_{forbid})|\varphi(T)\rangle\,,
\end{equation}
is depicted by dotted lines in fig.~\ref{fig:loss}a.
The smaller population transfer 
to the upper state in case of optimization with the state-dependent
constraint results in a larger 
robustness of the solution as the decay rate is increased. For a lifetime
on the order of the pulse duration, the final time objective is only
very slightly perturbed. In contrast the example without state-dependent
constraint  shows already a significant loss in the objective. This
effect becomes even 
more evident when the lifetime is smaller than the pulse duration. For
optimization without state-dependent constraint the transfer
efficiency is reduced by 50\% for lifetimes on the order of $100\,$fs.
For short lifetimes both optimization methods fail, nevertheless the algorithm
including the state-dependent constraint is still superior.
The decrease in efficiency of the transfer to the $v=1$ 
target level (solid lines) is associated
to the loss of population from the system (dotted lines)
as the lifetime decreases. 

An interesting effect is found as the lifetime becomes very small:
the final time objective actually improves. This is a manifestation of the 
quantum Zeno effect which states that a continuously observed quantum
system never decays \cite{QuZeno}. In our example, the usual
interpretation is inverted: The decay process 
can be associated with a weak measurement monitoring 
the population in the third electronic state, and
the optimized field corresponds to the decay in the usual picture. As
the decay rate increases, respectively the lifetime decreases, 
monitoring of the forbidden subspace becomes continuous and the pulse
cannot populate the lossy state anymore. The possibility to loose
population becomes thus smaller
(cf. dottes lines in  fig.~\ref{fig:loss}a) which
entails also better results for the final time objective (cf. solid
lines in  fig.~\ref{fig:loss}a). However, by 
comparing the solid and dotted lines in fig.~\ref{fig:loss}a, it is
obvious that the loss of coherence in the control process is larger than the
loss of population alone. 

Loss or decoherence is the main obstacle for generating controlled 
unitary transformations.
Figure \ref{fig:loss}b shows how the objective deteriorates due to loss
for the unitary transformation. The opimized field obtained 
with the state-dependent constraint is able to maintain a
high fidelity despite the loss term.
The performance of the  optimized field for the unitary transformation 
is clearly worse than for the state-to-state transition. This is due
to more population transfer to the upper electronic state in case of
the unitary transformation obtained after
the same number of iterations. A result comparable to that of
optimizing the state-to-state transition with the state-dependent
constraint can be achieved by increasing the number of iterations.


\section{Discussion}
\label{sec:disc}

The present work is related to a number of previous OCT studies using
state-dependent constraints, or equivalently, a time dependent target. 
The general formulation of our complete functional
for optimization of a state-to-state transition employing a
state-dependent constraint, Eq. (\ref{eq:functional_j}), is related to the functional  
of Ref. \cite{OhtsukiJCP04} identifying $\Op{D}=X$ and
$\Op{P}(t)=Y(t)$. Similarly, the specific application 
of Refs. \cite{KaiserJCP04,KaiserCP06} corresponds to
$J_0=0$ and $\Op{P}(t)=f(t)\Op{O}$; 
and the application of Ref. \cite{SerbanPRA05} to $J_0=0$, 
$\Op{P}(t)=\Op{O}_1(t)+2T\delta(t-T)\Op{O}_2$, and
$\Op{D}=\Op{O}_2$. The approaches of Refs. \cite{KaiserJCP04,KaiserCP06,SerbanPRA05} 
consist in imposing a specific desired dynamics onto
the system, for example, the populations following a
given time profile, or the maximization of the expectation value of a given operator.

In the studies of
Refs. \cite{OhtsukiJCP04,KaiserJCP04,KaiserCP06,SerbanPRA05}, the
optimization 
algorithm was obtained using the variational method. It is well-known
that the  variational approach is compatible with a large number of
implementations of the control algorithm, not all of them showing
monotonic convergence. Ref. \cite{OhtsukiJCP04} gives
a detailed analysis of this family. Our results which were obtained with the 
Krotov method, correspond to the case $\eta_k=0$ and $\xi_k=1$ of
Ref. \cite{OhtsukiJCP04}. The Krotov
method comes with the advantage of allowing a straightforward discussion
of the convergence of the algorithm in terms of the sign of the optimization
parameters $\lambda_i$. As discussed in Section \ref{sec:formalism}, this
sign of the optimization parameter turns out to be crucial for the choice 
of projector $\Op{P}(t)$.

As mentioned in the introduction, a different approach to avoid population
leakage to undesired states is using local control
theory \cite{k89,k94,Malinovsky97,Shlomo04}. The example chosen in Ref. \cite{Shlomo04}
differs from ours: In Ref. \cite{Shlomo04},
the allowed subspace was restricted to the register 
levels of a unitary transformation. This choice forces the dynamics of the system 
to rely solely on two-photon processes which assure that the population remains in
the register. A similar task in our formulation would correspond to choosing 
the operator $\Op{P}_{allow}$ as the projector onto the
X$^1\Sigma_g^+$ levels only.
However, the state-dependent constraint is formulated such as to
maximize the time-averaged population in the allowed subspace.
The family of solutions found by the 
algorithm consists in transferring some amount of population to
the intermediate electronic state, but only for a very short time.
This is reflected in the inset of fig.~\ref{fig:allow} in a sequence
of ``spikes'' in the population of the forbidden subspace as a 
function of time. Since the constraint of not populating any
electronically excited state is much 
more restrictive than the time-averaged formulation, the number of optimization steps
needed for reaching a specified efficiency and hence the complexity of
the solution increase largely  compared to the present results.

\section{Conclusions}
\label{sec:concl}

Steering a system to a desired objective has to always be balanced by the damage
induced by the steering process. In the present study optimal control theory was adopted
to include a positive objective to be maximized and a negative constraint to
be avoided. To ensure monotonic convergence the additional constraint
which depends on the state of the system was
incorporated in the Krotov method.

It turned out that the state-dependent constraint needs to be
formulated in terms of maximizing population in the allowed
subspace. While one could expect this to be 
equivalent to minimizing population in
the forbidden subspace, the resulting algorithms and
their convergence behavior differ markedly. 

The algorithm was applied to a simple model mimicking
vibrational manifolds in three electronic states of an Rb$_2$
molecule. Population transfer from the vibrational level
$v=0$ to $v=1$ of the electronic ground state and the implementation
of a Fourier transform in the levels $v=0,1,2,3$ were chosen as
optimization examples for a state-to-state transition and for a
unitary transformation. In both cases, the optimized fields induce
Raman-like transitions via an electronically excited
state. It was shown that optimization including the state-dependent
constraint indeed avoids population of a higher electronically excited
state. This state can either correspond to
a loss channel itself or represent a resonant intermediate state in an
unwanted multi-photon process. 

A similar task of promoting one objective while avoiding damage has been developed using 
local control theory \cite{k94}. It is important to note that the optimal solutions
are quite different. The local control scheme tries to minimize the
instantaneous population. 
Therefore it resorts to off-resonant two-photon transitions where
only a transient population exists. 
The OCT scheme minimizes the integrated population in the forbidden
subspace. 
As a result the solution can contain abrupt spikes of population. Since these spikes have
no time to go anywhere the system remains protected against damage. 

The success of the scheme was demonstrated in the ability to cope with
a real decay channel in the forbidden subspace.
The decay term causes loss of population and loss of phase or
decoherence. The state-to-state objective can  
be thought to work without long term coherence. Brief periods of coherence are sufficient
to generate the transitions. On the contrary the unitary operator objective has to
maintain phase coherence for the total period. The ability of the
algorithm to find solutions 
that can cope with this scenario is encouraging. The robustness of the
coherent control solution 
could result from a scheme based on a large number of interfering pathways.
In this case loss of a few pathways will only slightly hinder the final objective.

\begin{acknowledgments}
We would like to thank David Tannor for many fruitful discussions.
We gratefully acknowledge financial support from the Spanish MCT (FIS2004-05687,
FIS2007-64018) and the Gobierno de Canarias (PI2004/025), from
the Israel Science Foundation, and from 
the Deutsche Forschungsgemeinschaft (KO 2302/1-1).
The Fritz Haber
Center is supported by the Minerva Gesellschaft f\"{u}r die Forschung
GmbH M\"{u}nchen, Germany. 
\end{acknowledgments}

\appendix

\section{Review of  the Krotov method for optimal control theory}
\label{sec:krotov}

In the following the system Hamiltonian $\Op{H}$ is assumed to
be Hermitian. The derivation can easily be generalized to
non-Hermitian Hamiltonians \cite{OhtsukiJCP04}.
Let's define the states $|f\rangle$,
\begin{eqnarray}\label{eq:f}
|f\rangle&=&
-\frac{i}{\hbar}\Op{H}[\epsilon]|\varphi\rangle\,,\nonumber\\
\langle f|&=&
\langle\varphi|\frac{i}{\hbar}\Op{H}[\epsilon]\,,
\end{eqnarray}
which correspond to the total time derivative of $\varphi(t)$ in
Eq.~(\ref{eq:schrodinger}). 
As mentioned above, the rigorous utilization of the Krotov method 
is somewhat cumbersome and little instructive. For simplicity 
the case of a functional depending on two real functions denoted by
$\varphi$ and $\varphi^\dagger$ and a real control $\epsilon$ will be
presented. 
This simple case can be connected
to the original problem by the following relations,
\begin{eqnarray}\label{eq:correspondence}
\psi&\longleftrightarrow& |\psi\rangle\,,\nonumber\\
\psi^\dagger&\longleftrightarrow& \langle \psi|\,.
\end{eqnarray}
for $\psi=\varphi,\chi,f$. 
Our derivation follows closely  Ref.~\cite{SklarzPRA02},
but including the state-dependent constraint, i.e. the dependency on
$g_b$, cf. Eq.~\ref{eq:g_b}. 
The final equations for the original
problem, Eqs. (\ref{eq:chistate}-\ref{eq:eps1}), 
are obtained following the steps presented in this Appendix.
Moreover, our outline  can be employed to derive new optimization
schemes based on different choices of $J_0$, $g_a$ or $g_b$.

\subsection{The scalar function $\Phi$ and the functional $L$}

All functions considered in the derivation, 
$\varphi$, $\epsilon$, etc., depend on $t$ but for simplicity this
dependence will only be made explicit for the initial and 
the final time, e.g. $\varphi_0$, $\varphi_T$. The 
Krotov method is based on the introduction of the arbitrary scalar function 
$\Phi(t,\varphi,\varphi^\dagger)\,$, the functions
\begin{equation}
G(\varphi_T,\varphi_T^\dagger,\Phi_T)\,=\,J_0(\varphi_T,\varphi_T^\dagger)
\,+\,\Phi(T,\varphi_T,\varphi_T^\dagger)\,, 
\end{equation}
and
\begin{widetext}
\begin{eqnarray}\label{eq:R}
R(\varphi,\varphi^\dagger,\epsilon,\Phi) =
-g_a(\epsilon)-g_b(\varphi,\varphi^\dagger)\,+\,
\left[\frac{\partial\Phi}{\partial\varphi}\right] f 
\, + f^\dagger \left[\frac{\partial\Phi}{\partial\varphi^\dagger}\right] \,+\,
\frac{\partial\Phi}{\partial t}\,.
\end{eqnarray}
\end{widetext}
Note that
\begin{equation}
R=-(g_a+g_b)+\frac{d\Phi}{dt}\,,
\end{equation}
where it was used that $f$ ($f^\dagger$) is the total time derivative of
$\varphi$ ($\varphi^\dagger$).

A new functional can be defined
\begin{widetext}
  \begin{equation}
L[\varphi,\varphi^\dagger,\epsilon,\Phi]\,=\,
G(\varphi_T,\varphi_T^\dagger,\Phi_T)\,-\,
\Phi(0,\varphi_0,\varphi_0^\dagger)\,-\,
\int_0^T R(\varphi,\varphi^\dagger,\epsilon,\Phi)\,dt\,,
\end{equation}
\end{widetext}
with the interesting property 
\begin{equation}
L[\varphi,\varphi^\dagger,\epsilon,\Phi]\,=\,J[\varphi,\varphi^\dagger,\epsilon]\,.
\end{equation}
The Krotov method takes advantage of this property and the freedom in the choice of 
$\Phi$ to find an iterative algorithm that minimizes (maximizes) the original
functional $J$ \cite{SklarzPRA02,JosePRA03}. 

\subsection{The iterative algorithm}

Let's start with a given field $\epsilon^{(0)}$ and the corresponding
functions $\varphi^{(0)}$ and $\varphi^{\dagger(0)}$. 
The values of the  functionals are denoted by $J^{(0)}$ and $L^{(0)}$.
The functional $J$ ($L$) should be minimized
(maximization will be discussed later on).
The objective is then to determine a new control $\epsilon^{(1)}$, given the
functions $\varphi^{(1)}$ and $\varphi^{\dagger(1)}$, for which
\begin{equation}
J^{(0)}\,=\,L^{(0)}\,\geq\,J^{(1)}\,=\,L^{(1)} \,.
\end{equation}
For the time being, 
 the control $\epsilon$ and the functions $\varphi$,
 $\varphi^{\dagger}$ are treated as
``independent variables''. The Krotov method accomplishes the optimization
in two steps.
\begin{enumerate}
\item A function $\Phi$ is determined 
  such that $L[\varphi,\varphi^{\dagger},\epsilon,\Phi]$
  has a maximum for $\varphi^{(0)}$ y $\varphi^{\dagger(0)}$
  regardless of the control $\epsilon$. 
  The maximum condition implies second order (functional) derivatives
  of $L$ with respect to $\varphi$ and 
  $\varphi^{\dagger}$. Note that these derivatives should be evaluated ``at'' any field 
  $\epsilon$. However, the much simpler evaluation at $\epsilon^{(0)}$
  turns out to be sufficient \cite{SklarzPRA02}.
\item $\epsilon^{(1)}$ is determined such that the functional is
  minimized with respect to
  all possible controls $\epsilon$. This implies a second order (functional) 
  derivative of $L$ with respect to $\epsilon$.
  Since $\Phi$ is determined by the conditions of step 1, 
  this can be done without considering the effect of the change in
  the functional due to  
  change in $\varphi$ and $\varphi^\dagger$ induced by the new control.
\end{enumerate}
Finally, the necessary relation between the field and the states
through the evolution equation is imposed, resulting in 
\begin{equation}
L[\varphi^{(0)},\varphi^{\dagger(0)},\epsilon^{(0)},\Phi]\;\geq\;
L[\varphi^{(1)},\varphi^{\dagger(1)},\epsilon^{(1)},\Phi]\,.
\end{equation}

The conditions derived from the second order (functional) derivatives
are rather complicated \cite{SklarzPRA02}.
Fortunately, for the problems under consideration, the minimum and
maximum conditions can be relaxed to extremum conditions on the functional,
for which only first order (functional) derivatives are needed. Moreover,
the extremum conditions are common for
minimization and maximization of the functional, therefore both cases
are treated together below. Due to the relaxation of the minimum and
maximum to extremum conditions, 
an additional step is required: The monotonic convergence 
to the target value of the final algorithm must be checked.

\subsection{First step: Determining $\Phi$ up to first order}

The evaluation of an expression $[\dots]$ at $\epsilon^{(0)}$,
$\varphi^{(0)}$ and $\varphi^{\dagger(0)}$ is denoted by $[\dots]_{(0)}$.
The extremum in $L$ for $\varphi^{(0)}$ and $\varphi^{\dagger(0)}$ 
corresponds to the following conditions on $R$:
\begin{equation}\label{eq:extremeR}
\left[\frac{\partial R}{\partial \varphi}\right]_{(0)}\,=\,0\,,\;\;\;\;\;\;\;\;
\left[\frac{\partial R}{\partial \varphi^\dagger}\right]_{(0)}\,=\,0\,.
\end{equation}
For example, using Eq. (\ref{eq:R}), the second derivative reads
\begin{equation}
\left[\frac{\partial R}{\partial \varphi^\dagger}\right]\,=\,
-\frac{\partial g_b}{\partial \varphi^\dagger}\,+\,
\left[\frac{\partial\Phi}{\partial\varphi}\right]\,
\frac{\partial f}{\partial\varphi^\dagger}\,+\,
\frac{\partial f^\dagger}{\partial\varphi^\dagger}\,
\left[\frac{\partial\Phi}{\partial\varphi^\dagger}\right]\,+\,
\frac{d}{dt}\left[\frac{\partial\Phi}{\partial\varphi^\dagger}\right]\,,
\end{equation}
where we have used that 
\begin{equation}
\left[\frac{\partial}{\partial\varphi}\,
\frac{\partial\Phi}{\partial\varphi^\dagger}\right]f\,+\,
f^\dagger\left[\frac{\partial}{\partial\varphi^\dagger}\,
\frac{\partial\Phi}{\partial\varphi^\dagger}\right]\,+\,
\frac{\partial}{\partial t}\left[\frac{\partial\Phi}{\partial\varphi^\dagger}\right]\,=\,
\frac{d}{dt}\left[\frac{\partial\Phi}{\partial\varphi^\dagger}\right]\,.
\end{equation}
New functions $\chi$ and $\chi^\dagger$ are defined,
\begin{equation}
\chi(t)\,\equiv\,\left[\frac{\partial\Phi}{\partial\varphi^\dagger}\right]_{(0)}\,,
\;\;\;\;\;\;\;
\chi^\dagger(t)\,\equiv\,\left[\frac{\partial\Phi}{\partial\varphi}\right]_{(0)}\,.
\end{equation}
The second condition in Eq. (\ref{eq:extremeR}) becomes then
\begin{equation}\label{eq:chievolution}
\frac{d\chi}{dt}\,=\,-\chi^\dagger
\left[\frac{\partial f}{\partial\varphi^\dagger}\right]_{(0)}\,-\,
\left[\frac{\partial f^\dagger}{\partial\varphi^\dagger}\right]_{(0)}\chi
\,+\,\left[\frac{\partial g_b}{\partial\varphi^\dagger}\right]_{(0)}\,,
\end{equation}
and a similar result can be found for $\chi^\dagger$.
Moreover, the extremum condition on $L$ also implies,
\begin{equation}
\left[\frac{\partial G}{\partial \varphi_T}\right]_{(0)}\,=\,0\,,\;\;\;\;\;\;\;\;
\left[\frac{\partial G}{\partial \varphi_T^\dagger}\right]_{(0)}\,=\,0\,.
\end{equation}
For example the second derivative gives
\begin{equation}
\frac{\partial G}{\partial\varphi_T^\dagger}\,=\,
\frac{\partial J_0}{\partial\varphi_T^\dagger}\,+\,
\frac{\partial\Phi}{\partial\varphi_T^\dagger}\,.
\end{equation}
Using the previous definition of $\chi$, $\chi(T)$ can be identified,
\begin{equation}\label{eq:chicondition}
\chi(T)\,=\,\chi_T\,=\,
-\left[\frac{\partial J_0}{\partial\varphi_T^\dagger}\right]_{(0)}\,.
\end{equation}

Therefore the extremum conditions allow to determine the function
$\chi$ (and $\chi^\dagger$) 
using the evolution Eq. (\ref{eq:chievolution}) with the ``initial'' condition at time
$t=T$, Eq. (\ref{eq:chicondition}). With the knowledge of $\chi$ and
$\chi^\dagger$, the function $\Phi$ can be constructed  up to first
order,
\begin{equation}
\Phi_1[\,t,\varphi,\varphi^\dagger\,]\,=\,
\chi^\dagger(t)\,\varphi\,+\,\varphi^\dagger\,\chi(t)\,.
\end{equation}
The function $\Phi_1$ of the original problem is obtained
using Eq. (\ref{eq:correspondence}), 
\begin{eqnarray}
\Phi_1[\,t,\varphi,\varphi^\dagger\,] &=& \nonumber
\langle\chi(t)|\varphi(t)\rangle\,+\,\langle\varphi(t)|\chi(t)\rangle \\
&=&2{\rm Re}\left\{\langle\chi(t)|\varphi(t)\rangle\right\}\,.
\end{eqnarray}

\subsection{Second step: Determining $\epsilon^{(1)}$}

The extremum condition on $L$ with respect to the control,
evaluated in $\epsilon^{(1)}$ leads to
\begin{equation}
\left[\frac{\partial R}{\partial\epsilon}\right]_{(1)}\,=\,0\,,
\end{equation}
where $[\dots]_{(1)}$ denotes the evaluation of the expression $[\dots]$ 
at $\epsilon^{(1)}$, $\varphi^{(1)}$ and $\varphi^{\dagger(1)}$.
Using the definition of $R$, the derivative gives
\begin{equation}
\frac{\partial R}{\partial\epsilon}\,=\,
-\frac{\partial g_a}{\partial\epsilon}\,+\,
\left[\frac{\partial\Phi}{\partial\varphi}\right]\,
\frac{\partial f}{\partial\epsilon}\,+\,
\frac{\partial f^\dagger}{\partial\epsilon}\,
\left[\frac{\partial\Phi}{\partial\varphi^\dagger}\right]\,,
\end{equation}
where it was used that $\Phi$ does not explicitly depend on $\epsilon$.
Using $\Phi_1$, the previous equation results in,
\begin{equation}\label{eq:newfield}
-\left[\frac{\partial g_a}{\partial\epsilon}\right]_{(1)}\,+\,
\chi^\dagger\left[\frac{\partial f}{\partial\epsilon}\right]_{(1)}\,+\,
\left[\frac{\partial f^\dagger}{\partial\epsilon}\right]_{(1)}\chi\,=\,0\,.
\end{equation}
Since $g_a$ and the functions $f$ and $f^\dagger$, defined in Eq. (\ref{eq:f}), 
depend on the control, this equation allows to determine $\epsilon^{(1)}$.

\subsection{Conditions for monotonic convergence}

To check whether the iterative method shows monotonic convergence, we define
$\Delta_1$ and $\Delta_2(t)$,
\begin{widetext}
\begin{equation}
L[\varphi^{(0)},\varphi^{\dagger(0)},\epsilon^{(0)},\Phi_1^{(0)}]\,-\,
L[\varphi^{(1)},\varphi^{\dagger(1)},\epsilon^{(1)},\Phi_1^{(1)}]\,=\,
\Delta_1+\int_0^T\,\Delta_2(t)\,dt\,,
\end{equation}
\end{widetext}
where 
\begin{eqnarray}\label{eq:delta_1}
\Delta_1&=&G(\varphi_T^{(0)},\varphi_T^{\dagger(0)},\Phi_{1T}^{(0)})\,-\,
G(\varphi_T^{(1)},\varphi_T^{\dagger(1)},\Phi_{1T}^{(1)})\nonumber\\
&=&J_0^{(0)}-\Phi_{1T}^{(0)}-J_0^{(1)}+\Phi_{1T}^{(1)}\,,
\end{eqnarray}
and
\begin{equation}
\Delta_2(t)\,=\,R(\varphi^{(1)},\varphi^{\dagger(1)},\epsilon^{(1)},\Phi_1^{(1)})\,-\,
R(\varphi^{(0)},\varphi^{\dagger(0)},\epsilon^{(0)},\Phi_1^{(0)})\,.
\end{equation}
To simplify this expression, it is assumed that
\begin{equation}
\frac{\partial f^\dagger}{\partial\varphi}\,=\,0\,,\;\;\;\;\;
\frac{\partial f}{\partial\varphi^{\dagger}}\,=\,0\,,
\end{equation}
which is true for the original problem, and the function $R$ is split into
$R=R_a+R_b$,
\begin{widetext}
\begin{eqnarray}
R_a\,&=&\,-g_a\,-\,\chi^{*}\left[\frac{\partial f}{\partial\varphi}\right]_{(0)}\varphi
\,+\,\chi^{\dagger}f
\,-\,\varphi^{\dagger}\left[\frac{\partial f^{\dagger}}{\partial\varphi^{\dagger}}\right]_{(0)}\chi
\,+\,f^{\dagger}\chi\,, \\
R_b\,&=&\,-g_b\,+\,\left[\frac{\partial g_b}{\partial\varphi}\right]_{(0)}\varphi\,+\,
\varphi^{\dagger}\left[\frac{\partial g_b}{\partial\varphi^{\dagger}}\right]_{(0)}\,.
\end{eqnarray}
\end{widetext}
Therefore $\Delta_2(t)=\Delta_{2a}(t)\,+\,\Delta_{2b}(t)$ with
\begin{eqnarray}\label{eq:delta_2}
\Delta_{2a}(t)&=&\left[R_a\right]_{(1)}\,-\,\left[R_a\right]_{(0)}\,,\nonumber\\
\Delta_{2b}(t)&=&\left[R_b\right]_{(1)}\,-\,\left[R_b\right]_{(0)}\,.
\end{eqnarray}
If the objective is to minimize the functional $J$ ($J^{(0)}\geq J^{(1)}$), 
the sufficient but not necessary conditions for monotonic convergence are given by 
$\Delta_j\geq 0$, ($j=1,2,3$).
Analogously, in order to maximize the functional it is sufficient that
$\Delta_j\leq 0$, ($j=1,2,3$). Whether the $\Delta_j$ are positive or negative
is determined by the particular choice of $J_0$, $g_a$, and $g_b$. 
In Sec.~\ref{sec:formalism} the values of $\Delta$ are analyzed for the
cases under study. 

\subsection{The Krotov method for unitary transformations}

The previous approach can  easily be generalized to the case of unitary transformations.
The functional $J$ depends then on $2N_r$ real functions denoted by the set
$\{\varphi_n,\varphi_n^\dagger\}$. As a consequence, the scalar functions $\Phi$,
$G$ and $R$ will also depend on all of them. Moreover, the new dependence must be taken 
into account in the derivatives of the previous equations by the substitution
\begin{eqnarray}
\frac{\partial}{\partial\varphi}\,&\longrightarrow&\,
\sum_n\frac{\partial}{\partial\varphi_n}\,,  \nonumber \\
\frac{\partial}{\partial\varphi^\dagger}\,&\longrightarrow&\,
\sum_n\frac{\partial}{\partial\varphi_n^\dagger}\,.
\end{eqnarray}
The remaining procedure is analogous to the state-to-state case, 
and the relations (\ref{eq:correspondence}) can be used to obtain the
equations for the optimization algorithm given in Sec.~\ref{subsec:form_Utrafo}.



\begin{thebibliography}{31}
\expandafter\ifx\csname natexlab\endcsname\relax\def\natexlab#1{#1}\fi
\expandafter\ifx\csname bibnamefont\endcsname\relax
  \def\bibnamefont#1{#1}\fi
\expandafter\ifx\csname bibfnamefont\endcsname\relax
  \def\bibfnamefont#1{#1}\fi
\expandafter\ifx\csname citenamefont\endcsname\relax
  \def\citenamefont#1{#1}\fi
\expandafter\ifx\csname url\endcsname\relax
  \def\url#1{\texttt{#1}}\fi
\expandafter\ifx\csname urlprefix\endcsname\relax\def\urlprefix{URL }\fi
\providecommand{\bibinfo}[2]{#2}
\providecommand{\eprint}[2][]{\url{#2}}

\bibitem[{\citenamefont{Rice and Zhao}(2000)}]{RZ00}
\bibinfo{author}{\bibfnamefont{S.~A.} \bibnamefont{Rice}} \bibnamefont{and}
  \bibinfo{author}{\bibfnamefont{M.}~\bibnamefont{Zhao}},
  \emph{\bibinfo{title}{Optical control of molecular dynamics}}
  (\bibinfo{publisher}{John Wiley \& Sons}, \bibinfo{address}{New York},
  \bibinfo{year}{2000}).

\bibitem[{\citenamefont{Brumer and Shapiro}(2003)}]{SB03}
\bibinfo{author}{\bibfnamefont{P.}~\bibnamefont{Brumer}} \bibnamefont{and}
  \bibinfo{author}{\bibfnamefont{M.}~\bibnamefont{Shapiro}},
  \emph{\bibinfo{title}{Principles and Applications of the Quantum Control of
  Molecular Processes}} (\bibinfo{publisher}{Wiley Interscience},
  \bibinfo{year}{2003}).

\bibitem[{\citenamefont{Zhu et~al.}(1998)\citenamefont{Zhu, Botina, and
  Rabitz}}]{rabitz1}
\bibinfo{author}{\bibfnamefont{W.}~\bibnamefont{Zhu}},
  \bibinfo{author}{\bibfnamefont{J.}~\bibnamefont{Botina}}, \bibnamefont{and}
  \bibinfo{author}{\bibfnamefont{H.}~\bibnamefont{Rabitz}},
  \bibinfo{journal}{J. Chem. Phys.} \textbf{\bibinfo{volume}{108}},
  \bibinfo{pages}{1953} (\bibinfo{year}{1998}).

\bibitem[{\citenamefont{Kosloff et~al.}(1989)\citenamefont{Kosloff, Rice,
  Gaspard, Tersigni, and Tannor}}]{k67}
\bibinfo{author}{\bibfnamefont{R.}~\bibnamefont{Kosloff}},
  \bibinfo{author}{\bibfnamefont{S.~A.} \bibnamefont{Rice}},
  \bibinfo{author}{\bibfnamefont{P.}~\bibnamefont{Gaspard}},
  \bibinfo{author}{\bibfnamefont{S.}~\bibnamefont{Tersigni}}, \bibnamefont{and}
  \bibinfo{author}{\bibfnamefont{D.~J.} \bibnamefont{Tannor}},
  \bibinfo{journal}{Chem. Phys.} \textbf{\bibinfo{volume}{139}},
  \bibinfo{pages}{201} (\bibinfo{year}{1989}).

\bibitem[{\citenamefont{Sklarz and Tannor}(2002)}]{SklarzPRA02}
\bibinfo{author}{\bibfnamefont{S.~E.} \bibnamefont{Sklarz}} \bibnamefont{and}
  \bibinfo{author}{\bibfnamefont{D.~J.} \bibnamefont{Tannor}},
  \bibinfo{journal}{Phys. Rev. A} \textbf{\bibinfo{volume}{66}},
  \bibinfo{pages}{053619} (\bibinfo{year}{2002}).

\bibitem[{\citenamefont{Palao and Kosloff}(2003)}]{JosePRA03}
\bibinfo{author}{\bibfnamefont{J.~P.} \bibnamefont{Palao}} \bibnamefont{and}
  \bibinfo{author}{\bibfnamefont{R.}~\bibnamefont{Kosloff}},
  \bibinfo{journal}{Phys. Rev. A} \textbf{\bibinfo{volume}{68}},
  \bibinfo{pages}{062308} (\bibinfo{year}{2003}).

\bibitem[{\citenamefont{Gollub et~al.}(2008)\citenamefont{Gollub, Kowalewski,
  and de~Vivie-Riedle}}]{Gollub08}
\bibinfo{author}{\bibfnamefont{C.}~\bibnamefont{Gollub}},
  \bibinfo{author}{\bibfnamefont{M.}~\bibnamefont{Kowalewski}},
  \bibnamefont{and}
  \bibinfo{author}{\bibfnamefont{R.}~\bibnamefont{de~Vivie-Riedle}},
  \bibinfo{journal}{quant-ph/0801.3935}  (\bibinfo{year}{2008}).

\bibitem[{\citenamefont{Huang et~al.}(1983)\citenamefont{Huang, Tarn, and
  Clark}}]{clark}
\bibinfo{author}{\bibfnamefont{G.~M.} \bibnamefont{Huang}},
  \bibinfo{author}{\bibfnamefont{T.~J.} \bibnamefont{Tarn}}, \bibnamefont{and}
  \bibinfo{author}{\bibfnamefont{J.~W.} \bibnamefont{Clark}},
  \bibinfo{journal}{J. Math. Phys.} \textbf{\bibinfo{volume}{24}},
  \bibinfo{pages}{2608} (\bibinfo{year}{1983}).

\bibitem[{\citenamefont{Ramakrishna et~al.}(1995)\citenamefont{Ramakrishna,
  Salapaka, Dahleh, Rabitz, and Peirce}}]{rabitz2}
\bibinfo{author}{\bibfnamefont{V.}~\bibnamefont{Ramakrishna}},
  \bibinfo{author}{\bibfnamefont{M.~V.} \bibnamefont{Salapaka}},
  \bibinfo{author}{\bibfnamefont{M.}~\bibnamefont{Dahleh}},
  \bibinfo{author}{\bibfnamefont{H.}~\bibnamefont{Rabitz}}, \bibnamefont{and}
  \bibinfo{author}{\bibfnamefont{A.}~\bibnamefont{Peirce}},
  \bibinfo{journal}{Phys. Rev. A} \textbf{\bibinfo{volume}{51}},
  \bibinfo{pages}{960} (\bibinfo{year}{1995}).

\bibitem[{\citenamefont{Rabitz et~al.}(2006)\citenamefont{Rabitz, Ho, Hsieh,
  Kosut, and Demiralp}}]{rabitz3}
\bibinfo{author}{\bibfnamefont{H.}~\bibnamefont{Rabitz}},
  \bibinfo{author}{\bibfnamefont{T.-S.} \bibnamefont{Ho}},
  \bibinfo{author}{\bibfnamefont{M.}~\bibnamefont{Hsieh}},
  \bibinfo{author}{\bibfnamefont{R.}~\bibnamefont{Kosut}}, \bibnamefont{and}
  \bibinfo{author}{\bibfnamefont{M.}~\bibnamefont{Demiralp}},
  \bibinfo{journal}{Phys. Rev. A} \textbf{\bibinfo{volume}{74}},
  \bibinfo{pages}{012721} (\bibinfo{year}{2006}).

\bibitem[{\citenamefont{Bartana et~al.}(1997)\citenamefont{Bartana, Kosloff,
  and Tannor}}]{k131}
\bibinfo{author}{\bibfnamefont{A.}~\bibnamefont{Bartana}},
  \bibinfo{author}{\bibfnamefont{R.}~\bibnamefont{Kosloff}}, \bibnamefont{and}
  \bibinfo{author}{\bibfnamefont{D.~J.} \bibnamefont{Tannor}},
  \bibinfo{journal}{J. Chem. Phys.} \textbf{\bibinfo{volume}{106}},
  \bibinfo{pages}{1435} (\bibinfo{year}{1997}).

\bibitem[{\citenamefont{Bartana et~al.}(2001)\citenamefont{Bartana, Kosloff,
  and Tannor}}]{k163}
\bibinfo{author}{\bibfnamefont{A.}~\bibnamefont{Bartana}},
  \bibinfo{author}{\bibfnamefont{R.}~\bibnamefont{Kosloff}}, \bibnamefont{and}
  \bibinfo{author}{\bibfnamefont{D.~J.} \bibnamefont{Tannor}},
  \bibinfo{journal}{Chem. Phys.} \textbf{\bibinfo{volume}{267}},
  \bibinfo{pages}{195} (\bibinfo{year}{2001}).

\bibitem[{\citenamefont{Ohtsuki et~al.}(1999)\citenamefont{Ohtsuki, Zhu, and
  Rabitz}}]{otzki}
\bibinfo{author}{\bibfnamefont{Y.}~\bibnamefont{Ohtsuki}},
  \bibinfo{author}{\bibfnamefont{W.}~\bibnamefont{Zhu}}, \bibnamefont{and}
  \bibinfo{author}{\bibfnamefont{H.}~\bibnamefont{Rabitz}},
  \bibinfo{journal}{J. Chem. Phys.} \textbf{\bibinfo{volume}{110}},
  \bibinfo{pages}{9825} (\bibinfo{year}{1999}).

\bibitem[{\citenamefont{Palao and Kosloff}(2002)}]{JosePRL02}
\bibinfo{author}{\bibfnamefont{J.~P.} \bibnamefont{Palao}} \bibnamefont{and}
  \bibinfo{author}{\bibfnamefont{R.}~\bibnamefont{Kosloff}},
  \bibinfo{journal}{Phys. Rev. Lett.} \textbf{\bibinfo{volume}{89}},
  \bibinfo{pages}{188301} (\bibinfo{year}{2002}).

\bibitem[{\citenamefont{Tesch and de~Vivie-Riedle}(2002)}]{regina}
\bibinfo{author}{\bibfnamefont{C.}~\bibnamefont{Tesch}} \bibnamefont{and}
  \bibinfo{author}{\bibfnamefont{R.}~\bibnamefont{de~Vivie-Riedle}},
  \bibinfo{journal}{Phys. Rev. Lett.} \textbf{\bibinfo{volume}{89}},
  \bibinfo{pages}{157901} (\bibinfo{year}{2002}).

\bibitem[{\citenamefont{Rabitz et~al.}(2004)\citenamefont{Rabitz, Hsieh, and
  Rosenthal}}]{rabitz4}
\bibinfo{author}{\bibfnamefont{H.~A.} \bibnamefont{Rabitz}},
  \bibinfo{author}{\bibfnamefont{M.~M.} \bibnamefont{Hsieh}}, \bibnamefont{and}
  \bibinfo{author}{\bibfnamefont{C.~M.} \bibnamefont{Rosenthal}},
  \bibinfo{journal}{Science} \textbf{\bibinfo{volume}{303}},
  \bibinfo{pages}{1998} (\bibinfo{year}{2004}).

\bibitem[{\citenamefont{Kallush and Kosloff}(2006)}]{k218}
\bibinfo{author}{\bibfnamefont{S.}~\bibnamefont{Kallush}} \bibnamefont{and}
  \bibinfo{author}{\bibfnamefont{R.}~\bibnamefont{Kosloff}},
  \bibinfo{journal}{Phys. Rev. A} \textbf{\bibinfo{volume}{73}},
  \bibinfo{pages}{032324} (\bibinfo{year}{2006}).

\bibitem[{\citenamefont{Stefanatos et~al.}(2004)\citenamefont{Stefanatos,
  Khaneja, and Glaser}}]{glasser}
\bibinfo{author}{\bibfnamefont{D.}~\bibnamefont{Stefanatos}},
  \bibinfo{author}{\bibfnamefont{N.}~\bibnamefont{Khaneja}}, \bibnamefont{and}
  \bibinfo{author}{\bibfnamefont{S.~J.} \bibnamefont{Glaser}},
  \bibinfo{journal}{Phys. Rev. A} \textbf{\bibinfo{volume}{69}},
  \bibinfo{pages}{022319} (\bibinfo{year}{2004}).

\bibitem[{\citenamefont{Kosloff et~al.}(1992)\citenamefont{Kosloff, Hammerich,
  and Tannor}}]{k89}
\bibinfo{author}{\bibfnamefont{R.}~\bibnamefont{Kosloff}},
  \bibinfo{author}{\bibfnamefont{A.~D.} \bibnamefont{Hammerich}},
  \bibnamefont{and} \bibinfo{author}{\bibfnamefont{D.}~\bibnamefont{Tannor}},
  \bibinfo{journal}{Phys. Rev. Lett.} \textbf{\bibinfo{volume}{69}},
  \bibinfo{pages}{2172} (\bibinfo{year}{1992}).

\bibitem[{\citenamefont{Bartana et~al.}(1993)\citenamefont{Bartana, Kosloff,
  and Tannor}}]{k94}
\bibinfo{author}{\bibfnamefont{A.}~\bibnamefont{Bartana}},
  \bibinfo{author}{\bibfnamefont{R.}~\bibnamefont{Kosloff}}, \bibnamefont{and}
  \bibinfo{author}{\bibfnamefont{D.~J.} \bibnamefont{Tannor}},
  \bibinfo{journal}{J. Chem. Phys.} \textbf{\bibinfo{volume}{99}},
  \bibinfo{pages}{196} (\bibinfo{year}{1993}).

\bibitem[{\citenamefont{Malinovsky et~al.}(1997)\citenamefont{Malinovsky,
  Meier, and Tannor}}]{Malinovsky97}
\bibinfo{author}{\bibfnamefont{V.~S.} \bibnamefont{Malinovsky}},
  \bibinfo{author}{\bibfnamefont{C.}~\bibnamefont{Meier}}, \bibnamefont{and}
  \bibinfo{author}{\bibfnamefont{D.~J.} \bibnamefont{Tannor}},
  \bibinfo{journal}{Chem. Phys.} \textbf{\bibinfo{volume}{221}},
  \bibinfo{pages}{67} (\bibinfo{year}{1997}).

\bibitem[{\citenamefont{Sklarz and Tannor}(2004)}]{Shlomo04}
\bibinfo{author}{\bibfnamefont{S.~E.} \bibnamefont{Sklarz}} \bibnamefont{and}
  \bibinfo{author}{\bibfnamefont{D.~J.} \bibnamefont{Tannor}},
  \bibinfo{journal}{quant-ph/0404081}  (\bibinfo{year}{2004}).

\bibitem[{\citenamefont{Sklarz and Tannor}(2006)}]{Sklarz06}
\bibinfo{author}{\bibfnamefont{S.}~\bibnamefont{Sklarz}} \bibnamefont{and}
  \bibinfo{author}{\bibfnamefont{D.~J.} \bibnamefont{Tannor}},
  \bibinfo{journal}{Chem. Phys.} \textbf{\bibinfo{volume}{322}},
  \bibinfo{pages}{87} (\bibinfo{year}{2006}).

\bibitem[{\citenamefont{Ohtsuki et~al.}(2004)\citenamefont{Ohtsuki, Turinici,
  and Rabitz}}]{OhtsukiJCP04}
\bibinfo{author}{\bibfnamefont{Y.}~\bibnamefont{Ohtsuki}},
  \bibinfo{author}{\bibfnamefont{G.}~\bibnamefont{Turinici}}, \bibnamefont{and}
  \bibinfo{author}{\bibfnamefont{H.}~\bibnamefont{Rabitz}},
  \bibinfo{journal}{J. Chem. Phys.} \textbf{\bibinfo{volume}{120}},
  \bibinfo{pages}{5509} (\bibinfo{year}{2004}).

\bibitem[{\citenamefont{Kaiser and May}(2004)}]{KaiserJCP04}
\bibinfo{author}{\bibfnamefont{A.}~\bibnamefont{Kaiser}} \bibnamefont{and}
  \bibinfo{author}{\bibfnamefont{V.}~\bibnamefont{May}}, \bibinfo{journal}{J.
  Chem. Phys.} \textbf{\bibinfo{volume}{121}}, \bibinfo{pages}{2528}
  (\bibinfo{year}{2004}).

\bibitem[{\citenamefont{\c{S}erban et~al.}(2005)\citenamefont{\c{S}erban,
  Werschnik, and Gross}}]{SerbanPRA05}
\bibinfo{author}{\bibfnamefont{I.}~\bibnamefont{\c{S}erban}},
  \bibinfo{author}{\bibfnamefont{J.}~\bibnamefont{Werschnik}},
  \bibnamefont{and} \bibinfo{author}{\bibfnamefont{E.~K.~U.}
  \bibnamefont{Gross}}, \bibinfo{journal}{Phys. Rev. A}
  \textbf{\bibinfo{volume}{71}}, \bibinfo{pages}{053810}
  (\bibinfo{year}{2005}).

\bibitem[{\citenamefont{Kaiser and May}(2006)}]{KaiserCP06}
\bibinfo{author}{\bibfnamefont{A.}~\bibnamefont{Kaiser}} \bibnamefont{and}
  \bibinfo{author}{\bibfnamefont{V.}~\bibnamefont{May}},
  \bibinfo{journal}{Chem. Phys.} \textbf{\bibinfo{volume}{320}},
  \bibinfo{pages}{95} (\bibinfo{year}{2006}).

\bibitem[{\citenamefont{Koch et~al.}(2004)\citenamefont{Koch, Palao, Kosloff,
  and Masnou-Seeuws}}]{ChrPRA04}
\bibinfo{author}{\bibfnamefont{C.~P.} \bibnamefont{Koch}},
  \bibinfo{author}{\bibfnamefont{J.~P.} \bibnamefont{Palao}},
  \bibinfo{author}{\bibfnamefont{R.}~\bibnamefont{Kosloff}}, \bibnamefont{and}
  \bibinfo{author}{\bibfnamefont{F.}~\bibnamefont{Masnou-Seeuws}},
  \bibinfo{journal}{Phys. Rev. A} \textbf{\bibinfo{volume}{70}},
  \bibinfo{pages}{013402} (\bibinfo{year}{2004}).

\bibitem[{\citenamefont{Park et~al.}(2001)\citenamefont{Park, Suh, Lee, and
  Jeung}}]{ParkJMS01}
\bibinfo{author}{\bibfnamefont{S.~J.} \bibnamefont{Park}},
  \bibinfo{author}{\bibfnamefont{S.~W.} \bibnamefont{Suh}},
  \bibinfo{author}{\bibfnamefont{Y.~S.} \bibnamefont{Lee}}, \bibnamefont{and}
  \bibinfo{author}{\bibfnamefont{G.~H.} \bibnamefont{Jeung}},
  \bibinfo{journal}{J. Molec. Spec.} \textbf{\bibinfo{volume}{207}},
  \bibinfo{pages}{129} (\bibinfo{year}{2001}).

\bibitem[{\citenamefont{Weinstein et~al.}(2001)\citenamefont{Weinstein, Pravia,
  Fortunato, Lloyd, and Cory}}]{WeinsteinPRL01}
\bibinfo{author}{\bibfnamefont{Y.~S.} \bibnamefont{Weinstein}},
  \bibinfo{author}{\bibfnamefont{M.~A.} \bibnamefont{Pravia}},
  \bibinfo{author}{\bibfnamefont{E.~M.} \bibnamefont{Fortunato}},
  \bibinfo{author}{\bibfnamefont{S.}~\bibnamefont{Lloyd}}, \bibnamefont{and}
  \bibinfo{author}{\bibfnamefont{D.~G.} \bibnamefont{Cory}},
  \bibinfo{journal}{Phys. Rev. Lett.} \textbf{\bibinfo{volume}{86}},
  \bibinfo{pages}{1889} (\bibinfo{year}{2001}).

\bibitem[{\citenamefont{Misra and Sudarshan}(1977)}]{QuZeno}
\bibinfo{author}{\bibfnamefont{B.}~\bibnamefont{Misra}} \bibnamefont{and}
  \bibinfo{author}{\bibfnamefont{E.~C.~G.} \bibnamefont{Sudarshan}},
  \bibinfo{journal}{J. Math. Phys.} \textbf{\bibinfo{volume}{18}},
  \bibinfo{pages}{756} (\bibinfo{year}{1977}).

\end{thebibliography}
\end{document}